\documentclass[bbm,amsfonts,amsmath,amssymb,12pt]{article}
\usepackage{feynmf}
\usepackage[latin1]{inputenc}
\usepackage[T1]{fontenc}
\usepackage[english]{babel}
\usepackage{amssymb,amsmath}
\usepackage{amscd}
\usepackage{fancyhdr}
\usepackage{color}
\usepackage{hyperref}
\usepackage{epsfig}
\usepackage{graphicx}
\begin{document}

\newcommand{\drawsquare}[2]{\hbox{%
\rule{#2pt}{#1pt}\hskip-#2pt
\rule{#1pt}{#2pt}\hskip-#1pt
\rule[#1pt]{#1pt}{#2pt}}\rule[#1pt]{#2pt}{#2pt}\hskip-#2pt
\rule{#2pt}{#1pt}}

\newcommand{\Yfund}{\raisebox{-.5pt}{\drawsquare{6.5}{0.4}}}
\newcommand{\Yasymm}{\raisebox
{-3.5pt}{\drawsquare{6.5}{0.4}}\hskip-6.9pt%
                      \raisebox{3pt}{\drawsquare{6.5}{0.4}}%
                     }
\newcommand{\Ysymm}{\Yfund\hskip-0.4pt%
                     \Yfund}
\def\symm{\Ysymm}
\def\bsymm{\overline{\Ysymm}}
\def\drawbox#1#2{\hrule height#2pt
         \hbox{\vrule width#2pt height#1pt \kern#1pt
               \vrule width#2pt}
               \hrule height#2pt}

\def\Fund#1#2{\vcenter{\vbox{\drawbox{#1}{#2}}}}
\def\Asym#1#2{\vcenter{\vbox{\drawbox{#1}{#2}
               \kern-#2pt       
               \drawbox{#1}{#2}}}}
\def\sym#1#2{\vcenter{\hbox{ \drawbox{#1}{#2} \drawbox{#1}{#2}    }}}
\def\fund{\Fund{6.4}{0.3}}
\def\asymm{\Asym{6.4}{0.3}}
\def\bfund{\overline{\fund}}
\def\basymm{\overline{\asymm}}
\begin{titlepage}


\vspace{1cm}

\begin{center}

{\Large \bf Hidden QCD in Chiral Gauge Theories}
\end{center}

\vspace{0.5cm}

\begin{center}
{Thomas A. {\sc Ryttov}\footnote{E-mail: ryttov@nbi.dk} \quad and
\quad Francesco {\sc Sannino}\footnote{E-mail: sannino@nbi.dk}}
\end {center}
\begin{center}

{\it The Niels Bohr Institute, Blegdamsvej 17, Copenhagen \O,
Denmark.}\end{center}

\vspace{3mm}

\begin{abstract}
The 't Hooft and Corrigan-Ramond limits of massless one-flavor QCD
consider the two Weyl fermions to be respectively in the fundamental
representation or the two index antisymmetric representation of the
gauge group. We introduce a limit in which one of the two Weyl
fermions is in the fundamental representation and the other in the
two index antisymmetric representation of a generic SU(N) gauge
group. This theory is chiral and to avoid gauge anomalies a more
complicated chiral theory is needed. This is the generalized
Georgi-Glashow model with one vector like fermion.

We show that there is an interesting phase in which the considered
chiral gauge theory, for any N, Higgses via a bilinear condensate:
The gauge interactions break spontaneously to ordinary massless
one-flavor SU(3) QCD. The additional elementary fermionic matter is
uncharged under this SU(3) gauge theory. It is also seen that when
the number of colors reduce to three it is exactly this hidden QCD
which is revealed.
\end{abstract}

\end{titlepage}

\section{Introduction}

Different limits in gauge theories have been explored in the past
with the hope to get a better understanding of strongly interacting
gauge dynamics. The large $N$ limit a l\`a 't Hooft
\cite{'tHooft:1973jz} and its theoretical success
\cite{Witten:1979kh} have triggered a good deal of past and recent
work in different realms of theoretical physics from field theory to
string theory. Modern lattice simulations have recently explored
this limit \cite{Bringoltz:2005rr,Narayanan:2005gh} as well as
phenomenological studies based on low energy meson-meson scattering
amplitudes \cite{Harada:2003em}. One can, however, imagine different
limits which for certain values of the theory parameters yield
Quantum Chromo Dynamics (QCD). Here and in the following we focus
our attention on one-flavor QCD.

Immediately after the initial 't Hooft proposal Corrigan and Ramond
(CR) suggested a different limit \cite{Corrigan:1979xf} in which one
imagines the quarks to be in the antisymmetric rank-two
representation of the underlying gauge group\footnote{To be more
precise Corrigan and Ramond \cite{Corrigan:1979xf} suggested a
generalization of QCD in which some flavors were in the higher
dimensional representation and others still in the fundamental
representation. They focused the paper on the baryonic properties of
this theory and the $\eta^{\prime}$ properties at large $N$, as well
as technicolor models with fermions in the two index antisymmetric
representation of the gauge group. }. Clearly for three colors this
is the fundamental representation of the gauge group, however at
large number of colors the theory is not mapped in the 't Hooft
limit. For instance quark loops are suppressed in the 't Hooft limit
but not in the CR limit. There are a number of immediate
consequences suggested in the CR work and a decade later
investigated by Kiritsis and Papavassiliou \cite{Kiritsis:1989ge}.
Also very recently the idea that certain sectors of these theories
can be mapped in sectors of super Yang-Mills (SYM) has been
suggested \cite{Armoni:2003gp}. Using a supersymmetric inspired
effective Lagrangian approach some of the $1/N$ corrections were
investigated in \cite{hep-th/0309252}. When adding flavors the phase
diagram as a function of the number of flavors and colors has been
provided in \cite{hep-ph/0405209}. Further recent explorations of
this limit and its link to SYM can be found in \cite{hep-th/0507267}
while the reader can find a review of the string theory aspects in
\cite{DiVecchia:2004ct}.

The issue of confinement/deconfinement phase transition at nonzero
temperature for the CR limit has been explored in detail in
\cite{hep-th/0507251}. Here, one of the present authors has shown
the appearance of an alternating pattern as a function of number of
colors with respect to the remaining symmetries of the center group.
These results must be compared with the 't Hooft limit often used to
infer information about QCD.

New technicolor type theories have recently been constructed which
make use of new gauge interactions with fermions in the two index
(but symmetric) representation of the gauge group
\cite{hep-ph/0405209,hep-ph/0406200}. These theories are {\it not}
ruled out by the electroweak precision measurements
\cite{hep-ph/0505059}.

Since these two limits are very different it would be interesting to
explore a third one which is somewhat in between the 't Hooft and CR
one. To be more specific one would like to have half of the fermions
in the rank-two antisymmetric representation and the other half
transforming according to the fundamental representation of the
gauge group. If such a theory exists, for three colors it would be
identical to one-flavor QCD and for a generic number of colors the
theory would be, however, chiral. First we have certain issues to
address:

\begin{itemize}
\item
Can one construct a gauge anomaly free chiral theory which for three
colors is vector-like and in particular exactly one-flavor QCD ?
\item
Even if we find such a theory what can we say either about
QCD or this theory?
\end{itemize}

It turns out that it is possible to construct such a gauge anomaly
free chiral theory. It is a particular case of the generalized
Georgi-Glashow (gGG) model \cite{Georgi:1985hf}.
\begin{figure}[htbp]
\begin{center}
\includegraphics[scale=.5]{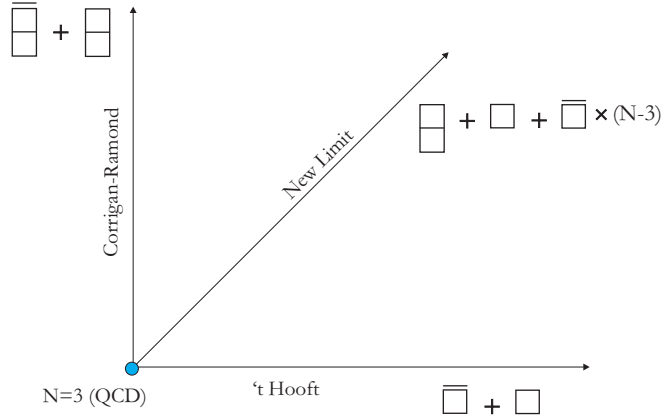}
\end{center}
\caption{Schematic representation of the matter fields required to
construct the three distinct limits which for three colors reduce to
massless one-flavor QCD. On each line we increase the number of
colors while the gauge group is always $SU(N)$.} \label{plot}
\end{figure}
In this model we have exactly one fermion in the two index
antisymmetric representation and other fermions in the fundamental
and antifundamental representation of the gauge group. Special
relations between the two kinds of fermions are required to avoid
the dangerous gauge anomalies. In Figure \ref{plot} we schematically
summarize the different limits for QCD. Together with their sister
theories of the Bars-Yankielowicz (BY) type \cite{Bars:1981se}, in
which the fermions with two indices are symmetrized, these chiral
gauge theories are known to display a number of interesting possible
phases in the infrared. See references
\cite{Peskin:1982mu,Appelquist:2000qg} for a review of some of these
properties and an exhaustive list of references on the gGG and BY
models.

In the next section we introduce the gGG model. We show the
existence of a phase in the model, for a generic number of colors,
in which the gauge group spontaneously breaks exactly to massless
one-flavor QCD and a remnant part which does not interact in the
deep infrared and is disconnected from QCD. This allows us to make a
number of predictions for the chiral gauge theory using our QCD
knowledge. We also discuss possible pitfalls and alternative
scenarios. We finally conclude in section three.

\section{The Generalized Georgi-Glashow Model: Hidden QCD}

The generalized Georgi-Glashow (gGG) model is based on the gauge
group $SU(N)$. The matter content of the theory includes fermions
$A=\psi^{[ij]}_{L},\ i,j=1,\cdots,N$ in the anti-symmetric
$\frac{1}{2}N(N-1)$ irreducible tensor representation; $N-4+p$
fermions $\bar{F}_{i,f}=\psi^{c}_{i,fL},\ f=1,\cdots,N-4+p$ in the
conjugated fundamental representation; and $p$ fermions
$F^{i,f}=\psi^{i,f}_{L},\ f=1,\cdots,p$ in the fundamental
representation as summarized in the table below:
\begin{center}
\begin{tabular}{c||ccccc }
& $ [SU(N)] $ & $ SU(N-4+p) $ & $ SU(p) $ & $ U_1(1) $ & $U_2(1)$ \\
\hline \hline \\
$ A $ & $\asymm$ & $1$ & $1$ & $N-4$ & $2p$ \\
&&&\\
$\bar{F}$ & $\bfund$ & $\bfund$ & $1$ & $-(N-2)$ & $-p$ \\
&&&\\
$F$ & $\fund$ & $1$ & $\fund$ & $N-2$ & $-(N-p)$   \\
\end{tabular}
\end{center}

The first $SU(N)$ is the gauge group. The two abelian symmetries are
anomaly free. They are linear combinations of the original $U(1)$
symmetries acting on each fermionic field separately. The
antifermion is needed to render the theory free from gauge
anomalies.

For now we just note that when $p=1$ and $N=3$ the theory coincides
with one-flavor QCD with $F$ being the quark and $A$ being the
antiquark.

The beta function of the model is
\begin{equation}
\beta = \mu \frac{d\alpha}{d\mu} = -
\beta_{1}(\frac{\alpha^2}{2\pi}) - \beta_2 (\frac{\alpha^3}{4\pi^2})
+ O(\alpha^4) \ ,
\end{equation}
where the terms of order $\alpha^4$ and higher are scheme-dependent
and $\beta_1 = 3N + 2-(2/3)p$ and $\beta_2 = (1/4)[13N^2 + 30N +1 -
12/N -2p((13/3)N-1/N)]$. Thus the theory is asymptotically free
if
\begin{equation} p < (9/2)N + 3 \ .
\end{equation}

There are a number of interesting possibilities studied in the past
for the ground state of these theories. The reader can find a list
and a guide on how to select among different possible ground states
in \cite{Appelquist:2000qg}. One of the remarkable properties of
these theories is that there are phases in which the whole chiral
symmetry does not break and the confining spectrum in the infrared
saturates all of the 't Hooft global anomalies of the theory.
Besides some of these phases can also be seen as Higgs phases or
partial Higgs phases and partial confining. By confining, however,
we simply mean that the low energy states are composites of the
underlying fermions in a color blind object. Since in these theories
the center group symmetry is absent we do not have a good order
parameter for the confinement/deconfinement transition. This is
similar to the case of one-flavor QCD.

As was noted above the case with $p=1$ is of particular interest.
For arbitrary $N$ this is an asymptotically free chiral theory which
for $N=3$, coincides with massless one-flavor QCD. It is also clear
that in the present limit to QCD the ordinary quark left emerges as
in the 't Hooft limit while the conjugated quark right is recovered
as in the CR limit. The fermions $\bar{F}$ disappear in the $N=3$
limit. In this sense we consider this theory as a novel limit
intermediate with respect to the 't Hooft and CR ones. A
straightforward count of the degrees of freedom in the ultraviolet
shows that we have $N^2$ gluons, $N^2$ $\bar{F}$-type fermions and
$N^2/2$ $A$-type fermions. On the other hand in the CR limit we have
$N^2$ gluons and $N^2$ fermions, while in the 't Hooft limit we just
have $N^2$ gluons. {}From the QCD point of view we suppress only
half of the fermion spectrum and keep the other half. $\bar{F}$ is a
fermion needed, for any finite $N$ to ensure that the theory is well
defined.

Although many phases are known and are possible to occur for a given
gGG theory still much is unknown about chiral gauge theories per se.
This heavily limits our predictive capabilities. However, it is
worth mentioning that a large $N$ limit for these theories can be
constructed and the expansion organized \cite{Eichten:1985fs}. So,
perhaps, also a string theory description of these theories is
possible.

We also notice that these theories have been used in
phenomenological models for grand unification, theories of massless
composite leptons and quarks and for suggesting purely fermionic
gauge theories which might naturally Higgs.

When departing from the three color limit we envision two logical
possibilities:
\begin{itemize}
\item
A phase transition in the number of colors occurs and the chiral
gauge theory in the new phase displays a vastly different structure
with respect to the three color limit.
\item
A more subtle phase emerges
which allows the theory to smoothly link itself to QCD when the
number of colors is three.
\end{itemize}

The first case may appear in different ways. Here we suggest the
following two possibilities. Assume that the gGG theory confines for
any number of colors without breaking any of the global symmetries.
Then the massless spectrum of this theory must be fermionic and
saturate all of the 't Hooft anomaly conditions. This solution does
exist and is summarized in the following table:
\begin{center}
\begin{tabular}{c||cccc }
& $ [SU(N)] $ & $ SU(N-3) $ &  $ U_1(1) $ & $U_2(1)$ \\
\hline \hline \\
$ \bar{F}A\bar{F} $ & $1$ & $\bsymm$ &  $-N$ & $0$ \\
&&&\\
$\bar{F}^{\dagger}A^{\dagger}F$ & $1$ & $\fund$ &  $N$ & $-N$ \\
\end{tabular}
\end{center}

Here the massless fermionic spectrum is constituted by two types of
colorless composite massless bound states, $ \bar{F}A\bar{F} $ and
$\bar{F}^{\dagger}A^{\dagger}F$. At large number of colors one can
drop the fermion $F$ and the only relevant bound state is
$\bar{F}A\bar{F}$. The phase transition from three colors to a
larger number of colors emerges since massless colorless bound
states appear which were not present before. Since the associated
flavor symmetries were not present in the QCD case one may still
argue that the associated currents simply do not match the ones of
QCD and that QCD is somewhat hidden in the massive dynamics of this
phase. Although we cannot completely exclude this possibility the
problem would, in any case, be that within this picture it is hard
to see any trace of QCD.

Another possible phase is the one in which the $F$ fermion condenses
with $\bar{F}$. In this case the gauge group remains intact while
the flavor symmetry breaks to $SU(N-4)\times U_1^{\prime}(1)\times
U_1^{\prime}(1)$ where the last two abelian symmetries are obtained
by combining the old abelian symmetries and one of the broken
nonabelian generators of the $SU(N-3)$ flavor symmetry. The
remaining theory is the one without the $F$ field and can either
Higgs or confine (these two cases lead to the same number of
massless fermions). Due to the breaking of the flavor symmetry we
also have goldstone bosons. The reader can find a review of this
phase in \cite{Appelquist:2000qg}. This phase is prohibited for $N$
less than four. We, hence, do not expect this phase to be smoothly
connected to the QCD limit.

We will now exhibit an interesting phase in which the theory
spontaneously breaks itself into massless one-flavor QCD and
elementary massless fermions uncharged under QCD. This can happen
for any number of colors. Recall that the full symmetry group of the
theory is
\begin{equation}
[SU(N)] \times SU(N-3) \times U_1(1) \times U_2(1) \ ,
\end{equation}
with the first one being the gauge symmetry. Lets consider the
following Higgs phase corresponding to the condensation in the
attractive channel
\begin{equation}
\asymm  \times \bfund \rightarrow \fund \ ,
\end{equation}
associated to the formation of the condensate
\begin{equation}
\langle A^{[ij]} \bar{F}_{j,f} \rangle \ ,
\end{equation}
where the spinor indices are contracted and $i,j = 1,\ldots,N$ and
$f=1,\ldots,N-3$. The present channel is more attractive than the
$F\bar{F}$ one when the colors are more than four in the one gluon
approximation of the potential between the two fermions. It breaks
the gauge symmetry from $SU(N)$ to $SU(3)$. Through color-flavor
locking the broken gauge subgroup $SU(N-3)$ combines with the flavor
group forming a new $SU'(N-3)$ global symmetry.

After spontaneous symmetry breaking we still have two abelian
symmetries $U_1'(1) \times U_2'(1)$ left intact. The first is formed
as a linear combination of the old abelian symmetries. The second is
formed by combining the broken diagonal $SU(N)$ generator
\begin{equation}
T_{(N\times N)} = \left(
\begin{array}{ccc|ccc}
3 &        &   &     &     & \\
  & \ddots &   &     &     & \\
  &        & 3 &     &     & \\
  \hline
  &        &   & 3-N &     & \\
  &        &   &     & 3-N & \\
  &        &   &     &     & 3-N
\end{array} \right) \ ,
\end{equation}
with the old $U_2(1)$ symmetry. Thus the pattern of symmetry
breaking is
\begin{equation}
\begin{array}{c}
[SU(N)] \times SU(N-3) \times U_1(1) \times U_2(1)
\\
\\
\downarrow \langle A\bar{F} \rangle
\\
\\
$[$SU(3)$]$ \times SU'(N-3) \times U'_1(1) \times U'_2(1) \ .
\end{array}
\end{equation}

Lets investigate the field content in the spontaneously broken
phase. All the Goldstone bosons are eaten up by the massive gauge
bosons. The fermions split into
\begin{equation}
A^{[ij]} = \left(
\begin{array}{cc}
A^{[ff']} & A^{[fc]} \\
A^{[cf]} & A^{[cc']}
\end{array}
\right)  \ ,
\end{equation}

\begin{equation}
\bar{F}_{i,f} = \left(
\begin{array}{c}
\bar{F}_{[f',f]} + \bar{F}_{\{f',f\}} \\
\bar{F}_{c,f}
\end{array} \right) \ ,
\end{equation}

\begin{equation}
F^{i} = \left(
\begin{array}{c}
F^{f} \\
F^{c}
\end{array} \right)  \ ,
\end{equation}
with $f,f'=1,\ldots,N-3$ and $c,c'=N-2,\ldots,N$. The massless
sector of the theory contains $F^c$ and $A^{[cc']}$ which belong to
the $\textbf{3}$ and $\bar{\textbf{3}}$ of the unbroken gauge
subgroup $SU(3)$. We identify them with an ordinary quark and
anti-quark respectively. Also $F^f \equiv L^f$ and
$\bar{F}_{\{f',f\}}\equiv \bar{L}_{\{f',f\}}$ which are color
singlets remain massless. Summarizing, the spontaneously broken
phase has the following massless fermion content
\begin{center}
\begin{tabular}{c||cccc}
& $ [SU(3)] $ & $ SU'(N-3) $  & $ U'_1(1) $ & $U'_2(1)$ \\
\hline \hline \\
$ q $ & $\fund$ & $1$ & $-N$ & $\frac{2}{3}N$ \\
&&&\\
$\bar{q}$ & $\bfund$ & $1$ & $N$ & $-\frac{2}{3}N$ \\
&&&\\
$L$ & $1$ & $\fund$ & $-N$ & $N$   \\
&&&\\
$\bar{L}$ & $1$ & $\bsymm$ & $-N$ & $0$
\end{tabular}
\end{center}

This shows that for any number of colors we have found a phase of
the theory which at low energy splits into massless one-flavor QCD
and a number of elementary massless fermions uncharged under the
color gauge group. However massless one-flavor QCD confines and gaps
as well so that in the deep infrared the only true massless degrees
of freedom are indeed the chiral fermions $L$ and $\bar{L}$. Since
the global symmetries are unbroken the massless spectrum in the deep
infrared, needed to saturate the 't Hooft anomalous conditions,
coincides with the one of the confining phase. However the massless
spectrum of the theory consists of elementary fermions as opposed to
the confining phase which consists of bound states of the original
elementary fermions. To observe the physical difference between
these different phases one has to go beyond the simple counting of
massless degrees of freedom. For instance one should probe the
structure of the massless fermions and investigate the massive
spectrum of the theory.

In our case we predict different energy ranges: First an energy
range above and around the scale $\Lambda$ of the condensate
$\langle A\bar{F}\rangle$. Some of the underlying fermions (which
can be traced via the analysis performed above) receive mass via
this condensate with a mass of the order of $\Lambda$. Except for
the massless octet of ordinary gluons all of the remaining gauge
bosons will acquire a mass of the order of $g\Lambda \lesssim
\Lambda$ where $g$ is the coupling constant evaluated at the energy
$\Lambda$. A second energy range is below the scale $\Lambda$ where
we have massless one-flavor QCD together with the $L$ and $\bar{L}$
fermions which are uncharged under the QCD gauge group. Then a new
scale $\Lambda_{QCD}<\Lambda$ develops where QCD confines. The third
and last energy range is below $\Lambda_{QCD}$ where only $L$ and
$\bar{L}$ survive as massless elementary fermions.

It is very tempting to assume a smooth limit as function of the
number of colors near the three color case. If this happens we can
predict that the QCD spectrum describes a good part of the massive
non-perturbative spectrum of the underlying chiral gauge theory.
This is so since the Lagrangian splits into two sectors below the
scale $\Lambda$ as follows
\begin{eqnarray}
\mathcal{L}=\mathcal{L}_{QCD}(q,\bar{q}) +
\mathcal{L}_{Non-Gauge}(L,\bar{L}) \ . \quad {\rm
At~energy~scales~smaller~than~\Lambda} \ .
\end{eqnarray}

Hence the partition function of the underlying chiral gauge theory
factorizes at low energy in the partition function for QCD and the
one of low energy massless noninteracting fermions. A part of the
massive spectrum of the gGG model corresponds to the one of
one-flavor QCD for any number of colors. Surprisingly this phase is
now smoothly connected to pure massless one-flavor QCD when the
number of colors is three even though our theory is a complicated
chiral gauge theory.

When the number of colors is very large the $A$ and $\bar{F}$
fermions dominate the action. This is so since the number of these
fermions increases quadratically with the number of colors while the
$F$ fermions increase only linearly with respect to the number of
colors. Since QCD emerges, in our picture, as a mismatch between the
$N$ colors and the $N-3$ flavors we expect this phenomenon to appear
only when considering subleading terms in the $1/N$ expansion of the
full theory.

\section{Conclusions}

We have introduced a new limit for massless one-flavor QCD. {}For
three colors one can imagine one of the two Weyl fermions of
one-flavor massless QCD to belong to the fundamental representation
of the gauge group and the other in the two index antisymmetric
representation. {}When the number of colors is larger than three the
theory is chiral. To avoid gauge anomalies a more complicated theory
is needed. This is the generalized Georgi-Glashow model with one
vector like fermion. While a large number of possible phases for
these theories have been constructed, unfortunately very little is
known about the ground state of these type of gauge theories and/or
the spectrum of massive states. It is possible that when the number
of colors increases a possible link with QCD is completely lost. We
have shown how this could happen by providing different phases in
which this theory would not display any QCD property.

Somewhat surprisingly, we have discovered a phase in which, for any
$N$, if we assume the $A\bar{F}$ bilinear to condense, (here we are
considering the case of $p=1$) the gauge interaction breaks
spontaneously to ordinary massless one-flavor $SU(3)$ QCD. The
additional elementary fermionic matter is uncharged under this
$SU(3)$ gauge theory. We have already stressed that such a
condensation pattern is partially supported by the fact that the
condensate corresponds to a maximally attractive channel when the
number of colors is sufficiently large. Besides, when the number of
colors reduce to three it is exactly this hidden QCD which is
revealed. The two sectors of the theory are disconnected at low
energy and we predict the one-flavor QCD spectrum to constitute a
relevant part of this certain class of generalized Georgi-Glashow
models. The large number of colors limit enhances only half of the
QCD fermionic matter. We expect that our limit can be used also when
exploring supersymmetric chiral gauge theories.

Together with the 't Hooft and Corrigan-Ramond limits, the present
one exhausts all of the possible ways (representation wise) one can
envision extending massless one-flavor QCD.

\vskip 2cm \centerline{\bf Acknowledgments} We thank S. Bolognesi,
D.D. Dietrich, P. Di Vecchia, M. Frandsen, C. Kouvaris, R. Marotta,
M. Shifman, B. Svetitsky and K. Tuominen for discussions and
comments.

\noindent TR wishes to thank the Lørup Fund and A.D. Jackson for
financial support during his stay at CERN. FS is supported by the
Marie Curie Excellence Grant as team leader under contract
MEXT-CT-2004-013510 and by the Danish Research Agency.  We also
thank the CERN theory division for the kind hospitality during the
initial stages of this work.

\end{document}